\documentclass{elsart}
\usepackage{psfig}
\usepackage{natbib}

\begin{document}

\runauthor{Wills, Shang and Yuan}


\begin{frontmatter}

\title{H$\beta$ Line Width and the UV-X-ray Spectra of Luminous AGN}

\author{Beverley J. Wills, Zhaohui Shang \& Juntao M. Yuan}
\address{Department of Astronomy, RLM 15.308,
The University of Texas, Austin, Texas 78712 USA}

\begin{abstract}
The width of the broad H$\beta$ emission line is the primary defining
characteristic of the NLS1 class.  This parameter is also an important
component of Boroson and Green's optical ``Eigenvector 1''(EV1), which
links steeper soft X-ray spectra with narrower H$\beta$ emission,
stronger H$\beta$ blue wing, stronger optical Fe\,II emission, and weaker
[O\,III]\,$\lambda$5007. Potentially, EV1 represents a fundamental
physical process linking the dynamics of fueling and outflow with the
accretion rate. We attempted to understand these relationships by
extending the optical spectra into the UV for a sample of 22 QSOs with
high quality soft-X-ray spectra, and discovered a whole new set of UV
relationships that suggest that high accretion rates are linked to dense
gas and perhaps nuclear starbursts. While it has been argued that narrow
(BLR) H$\beta$ means low Black Hole mass in luminous NLS1s, the
C\,IV\,$\lambda$1549 and Ly$\alpha$ emission lines are broader, perhaps
the result of outflows driven by their high Eddington accretion rates. We
present some new trends of optical-UV with X-ray spectral energy
distributions. Steeper X-ray spectra appear associated with stronger UV
relative to optical continua, but the presence of strong UV absorption
lines is associated with depressed soft X-rays and redder optical-UV
continua.

\end{abstract}


\end{frontmatter}


\section{Introduction}

The Boroson and Green (1992, BG92) ``Eigenvector 1'' (EV1) is one of the
reasons NLS1 are interesting, and why we are having a meeting about them.
EV1 (also called Principal Component 1) is a linear combination of
correlated optical and X-ray properties representing the greatest
variation among a set of spectra.  EV1 links narrower (BLR) H$\beta$ with
stronger H$\beta$ blue wing, stronger Fe\,II optical emission, and weaker
[OIII]\,$\lambda$5007 emission from the NLR, with steeper soft X-ray
spectra (Laor et al. 1994, 1997). The narrower H$\beta$ defining NLS1s
has been suggested to result from lower Black Hole masses, and therefore higher
accretion rates relative to the Eddington limit in these luminous AGN.
The NLS1s' steep X-ray spectra are also suggested to tie in with high Eddington
accretion rates (e.g. Laor, these proceedings).

In order to understand these relationships, we have extended the optical
spectra of an essentially complete sample of 22 PG QSOs to UV
wavelengths.  The QSOs of our sample are selected (by Laor et al. 1994)
to have low Galactic hydrogen absorption in order to derive reliable soft
X-ray slopes and intrinsic absorption.  The uncertainties in absorption
correspond to less than 5\% uncertainty in flux density at Ly$\alpha$.
Five of these QSOs are NLQ1s\footnote {At optical wavelengths an
observational distinction has been made between the higher luminosity
QSOs (L $> 10^{11.3}$ L$_\odot$, H$_0 = 100$ km s$^{-1}$ Mpc$^{-1}$),
where nuclear light dominates that from the host galaxy, and the lower
luminosity Seyfert galaxies.  We refer to QSOs with narrow BLR H$\beta$
as NLQ1s. } (with H$\beta$ FWHMs $< 2000$ km s$^{-1}$): PG 1001+054 (a
BAL QSO, 1740 km s$^{-1}$), PG 1115+407 (1720 km s$^{-1}$), PG 1402+261
(1910 km s$^{-1}$), PG 1440+356 (Mrk 478, 1450 km s$^{-1}$) and PG
1543+489 (1560 km s$^{-1}$). For comparison with X-ray selected samples,
we note that the $U-B$ color criterion for selecting the PG QSOs (Schmidt
\& Green 1983) is roughly equivalent to the hardness-ratio
criterion (HR1 $<$0) used to select soft X-ray AGNs (e.g. Grupe et al.
1998).

We address different aspects of the Eigenvector 1 relationships:
(i) The correlations among UV and optical emission lines, and soft X-ray slope.
We also present the Eigenvector 1 UV-optical relationships via a
spectral principal component analysis.
(ii) We show relationships among line widths, and
(iii) we show new relationships between the UV-optical and X-ray spectral
energy distributions (SEDs). 
We show how the 5 NLQ1s of our sample fit into these relationships.

\section{The UV Eigenvector 1 Relationships}

\begin{figure}[htb]
\centerline{\psfig{figure=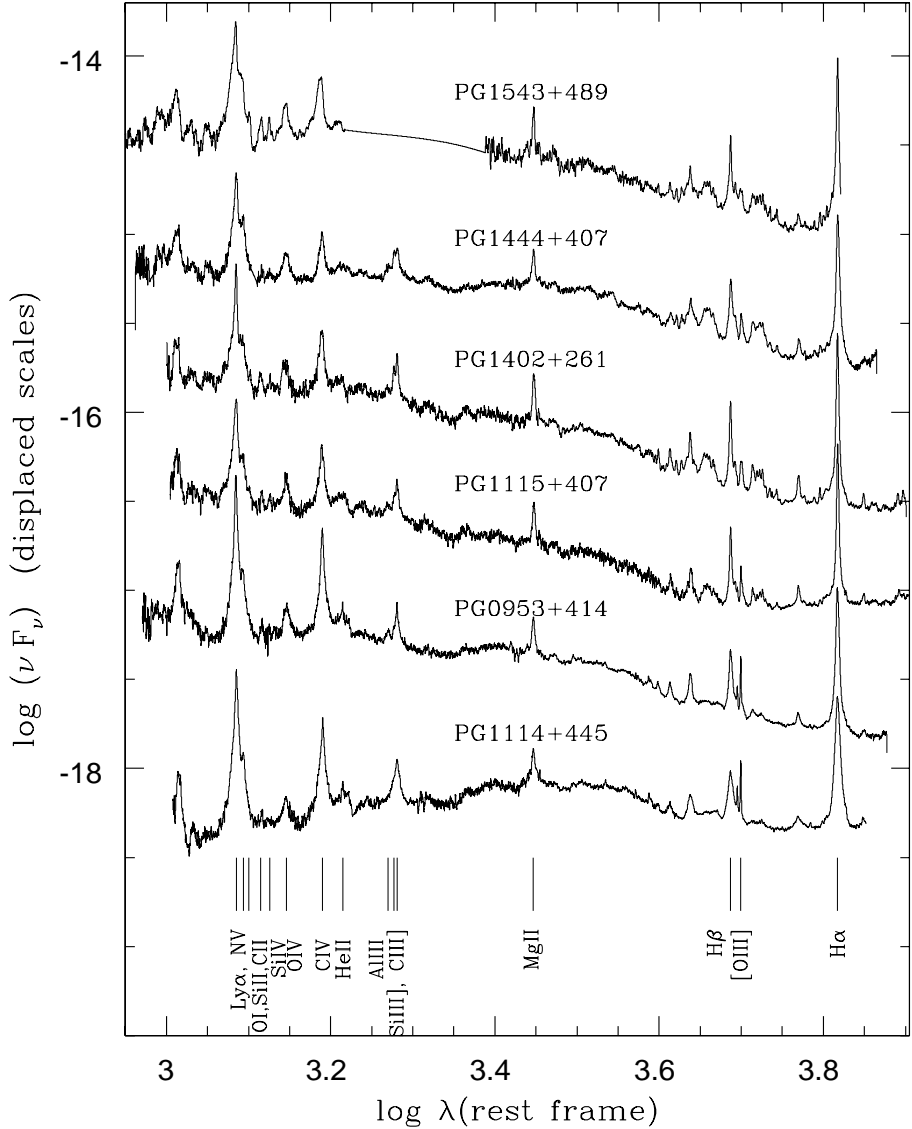,height=15truecm,width=12truecm,angle=0}}
\caption{
Above, Hubble Space Telescope (UV, Faint Object Spectrograph) and McDonald
Observatory (ground-based) spectra for several of the objects in the
complete sample.  These are in order of optical-X-ray Principal Component
1 (BG92, Laor et al. 1994), with the strongest Fe\,II(optical) and
weakest [O\,III] NLR emission at the top. }
\end{figure}

\begin{figure}[htb]
\centerline{\psfig{figure=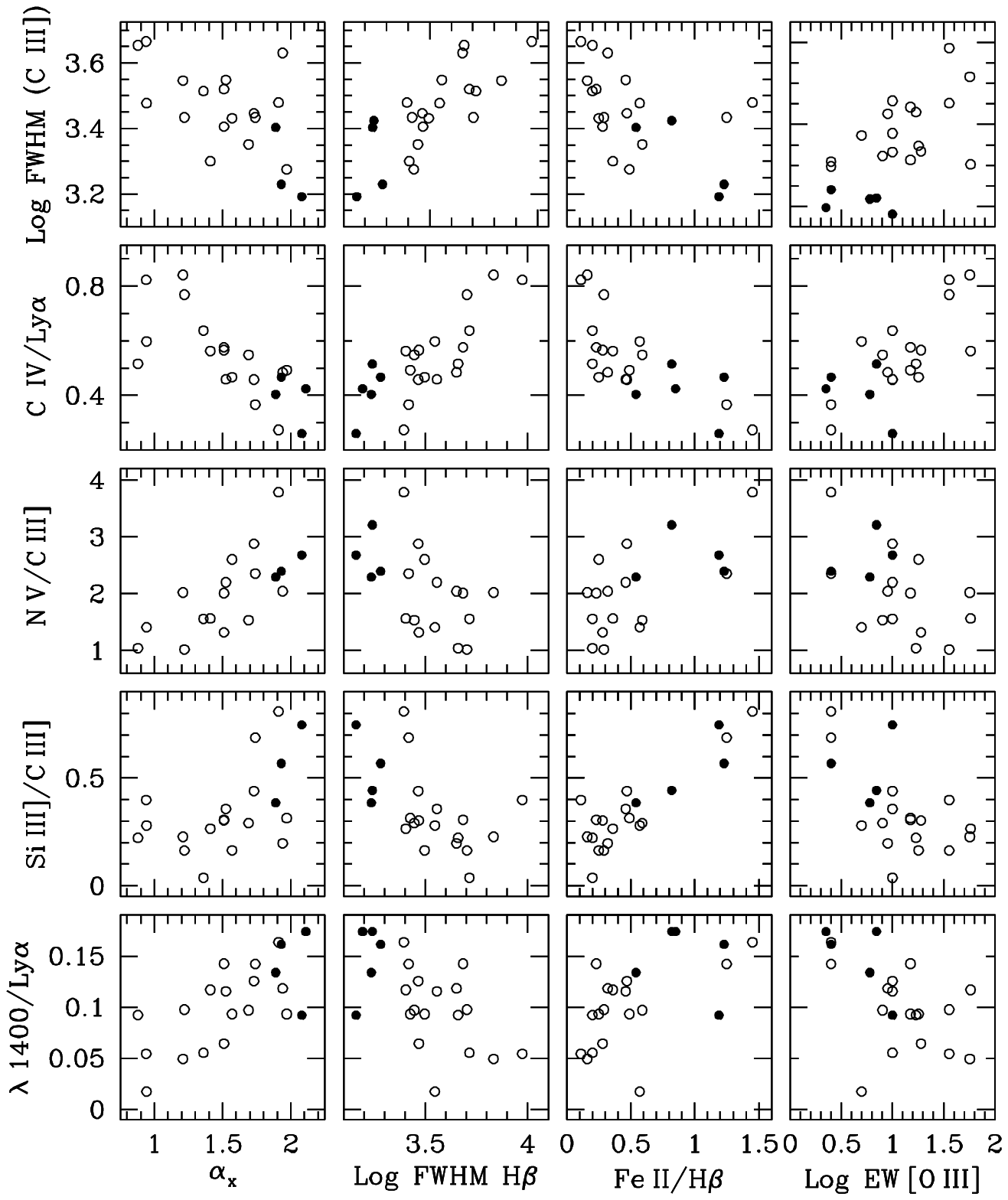,height=15truecm,width=13.0truecm,angle=0}}
\caption{
Correlations corresponding to the variations illustrated in Fig.\,1. New
UV observables are displayed vs. $\alpha_x$ and some optical (BG92)
Principal Component 1 parameters.  The two-tailed probability of these
correlations arising by chance from unrelated variables is between 1 in
50 and 1 in $>>1000$. }
\end{figure}

Examples of our spectra, over the entire UV-optical range, presented in
order of optical EV1, are shown in Fig.\,1 (see also Wills et al.
1999a,b, Francis \& Wills 1999). Inspection shows that, with changing
EV1, in the sense of decreasing H$\beta$ width and increasing strength of
the Fe\,II(optical) blends, the UV spectrum changes: C\,IV\,$\lambda$1549
is weaker and broader, Si\,III]\,$\lambda$1892 and nearby Fe\,III blends
become more prominent relative to C\,III]\,$\lambda$1909, and the
strength of low-ionization O\,I\,$\lambda$1304 and C\,II\,$\lambda$1335
increases.  Low velocity C\,IV gas appears to be strongly related to the
optical NLR [O\,III]\,$\lambda$5007 emission (Brotherton \& Francis
1999). Because Si\,III]\,$\lambda$1892 has a higher critical density than
the C\,III] transition, we surmise that increasing Fe\,II(optical), and
narrower H$\beta$, correspond to the increasing contribution from very
dense gas, $> 10^{10.5}$ cm$^{-3}$.  Unexpectedly, the high-ionization
N\,V\,$\lambda$1240 line strength appears to increase with increasing
density. We speculate that the higher-density gas is nitrogen enriched,
and the copious dense gas is a result of circumnuclear starbursts. As
suggested by BG92, this dense gas reduces the flux of ionizing photons
reaching the more extended NLR, resulting in the inverse
Fe\,II(optical)--[O\,III] relationship.  In Fig.\,1 the continuum of
PG\,1114+445 is seen to be significantly reddened; apparently dust is
associated with the high-ionization UV line absorption and the X-ray warm
absorption in this QSO (George et al. 1997; Mathur et al. 1998).

Some correlations between optical EV1 parameters and new UV emission-line
parameters are illustrated in Fig.\,2.  The NLQ1s are distinguished by
filled circles.  The two-tailed probability of these correlation arising
from truly unrelated variables ranges from 1 in 50 to 1 in $>>1000$.
Note that NLQ1s also contribute to these trends.  In the UV they have
weaker, broader C\,IV, stronger N\,V, stronger low-ionization lines and
more dense gas (as indicated by a larger ratio,
Si\,III]\,$\lambda$1892/C\,III]\,$\lambda$1909). Note also that
correlations with the X-ray spectral index $\alpha_x$ exist despite the
characteristic rapid X-ray variability of the NLQ1s.

\subsection{Spectral Principal Component Analyses}

\begin{figure}[htb]
\centerline{\psfig{figure=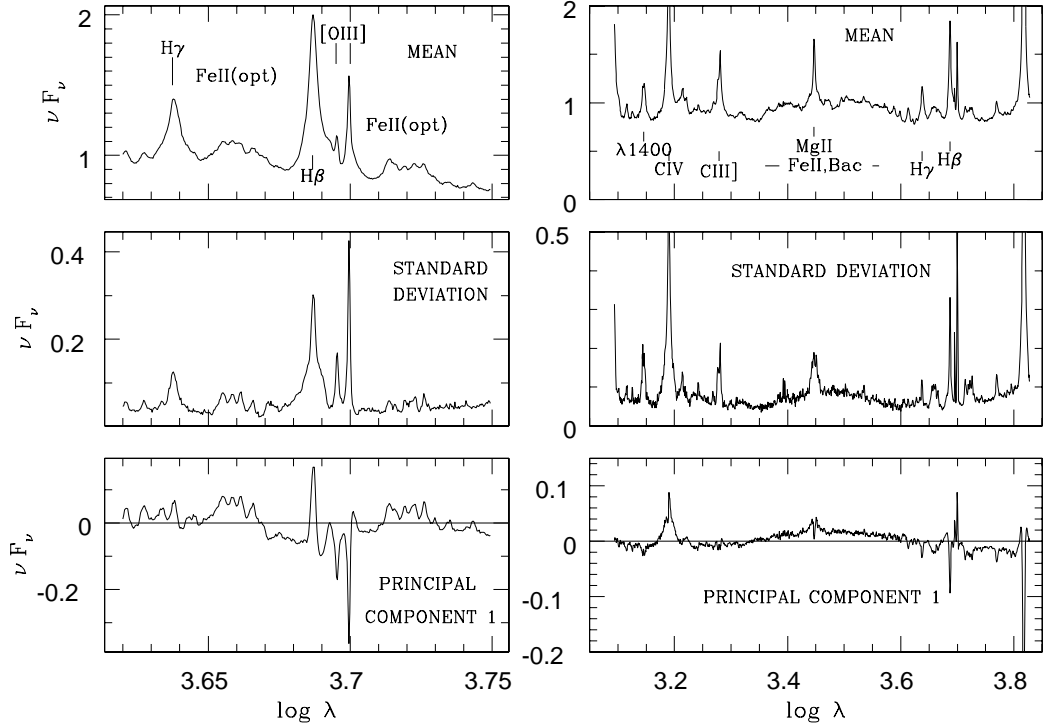,height=9.58truecm,width=13.8truecm,angle=0}}
\caption{
The aligned mean, standard deviation and principal component 1 spectra in
the H$\beta$ region (left) and the C\,IV\,$\lambda$1549 -- H$\alpha$
region (right). Emission from the lowest velocity gas, in the Narrow and
Broad Line Regions (NLR, BLR) accounts for most of the
spectrum-to-spectrum diversity. }
\end{figure}

The above trends can be illustrated in another informative way. Fig.\,3
shows the mean spectrum in the H$\beta$ region (left) and in the range
from C\,IV\,$\lambda$1549 to H$\alpha$ (right).  Ly$\alpha$ is excluded
to avoid the complication of strong absorption in a few objects. A
standard deviation spectrum in the middle panel shows that most of the
spectrum-to-spectrum variation appears to arise from emission from
low-velocity gas.  This is true for Ly$\alpha$, CIV, the Balmer lines,
and Fe\,II(optical) blends.  Because the sample is small we show only the
first eigenvector, approximately equivalent to BG92's EV1 (lower panel).
Now we can see which parts of the spectrum vary together or are
anticorrelated. The well-known BG92 correlation between the blended
Fe\,II(optical) emission and the narrow BLR H$\beta$ shows clearly, as
well as their anticorrelation with narrow
[O\,III]\,$\lambda\lambda$4959,5007.  Detailed inspection
of the H$\beta$ region shows
that the lower-velocity gas of the Fe\,II(optical) emission contributes
most to the correlations.  (Compare the resolved peaks in the middle and
lower panels with the smoother Fe\,II(optical) blends in the mean
spectrum). The UV Fe\,II blends between about 2200\AA\ and 3600\AA\
(log\,$\lambda$ between 3.34 and 3.56) appear anticorrelated with the
strength of optical Fe\,II blends -- probably an effect of optical depth
in the optically-thick UV resonance transitions, giving rise to the lower
optical depth Fe\,II optical transitions. This is consistent with the
explanation for the weaker [O\,III] in objects with strong
Fe\,II(optical) emission.

This simple interpretation is preliminary but serves to relate the
significant correlations with the overall spectrum.  Spectral principal
component analysis is a powerful tool for analysing linear correlations.
Possibly NLS1-type gas is actually present in all QSOs, but
with differing contributions to the overall spectra.  If so, this would
represent a different point of view on the significance of NLS1 line
widths. One point remains clear -- the luminous NLS1s are,
spectroscopically, simply more extreme than broader-line QSOs, but not
fundamentally different.

\section{Line Widths}

\begin{figure}[htb]
\centerline{\psfig{figure=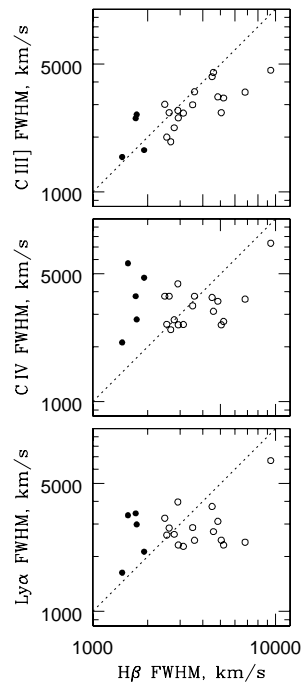,height=9truecm,width=3.71truecm,angle=0}}
\caption{Linewidth relationships.  The C\,III]\,$\lambda$1909 line has been
deblended from Si\,III]\,$\lambda$1892 and, where significant, from Fe\,III 
emission.  The C\,IV FWHM is as observed, not corrected for doublet separation,
which would be less than a 10\% correction.  Ly\,$\alpha$ has been
deblended from N\,V\,$\lambda$1240.  Note the significant ($< 2 \times 10^{-5}$
probability of it arising by chance) correlation between FWHM of C\,III] and
H$\beta$, but not for C\,IV or Ly$\alpha$.  The dashed lines indicate equal
line-width relationships.
 }
\end{figure}

In a sample with a wide range in luminosity, line width is seen to increase
with luminosity.  This leads one to question whether the defining criterion for
NLS1s or NLQ1s should be a function of luminosity!
Even for our small sample the trends of FWHM with luminosity
are just significant at the $\sim 1$\% level.   The data are consistent with a 
(luminosity)$^{1/4}$ dependence of linewidth expected if gravity dominates the
cloud motions (e.g. Laor, these proceedings). 
Anyway, within our sample, EV1 dominates the spectrum-to-spectrum variations.

Correlations between the widths of the UV broad lines and that of H$\beta$
are shown in Fig.\,4.  The widths of C\,III]\,$\lambda$1909 and H$\beta$
are strongly correlated, and similarly within the same QSO spectrum.  The
lower-ionization C\,III] emission probably arises in the same region as
H$\beta$. It seems surprising that the widths of the hydrogen lines,
H$\beta$ and Ly$\alpha$, are not correlated.  Neither Ly$\alpha$ or C\,IV
widths are correlated with C\,III] or H$\beta$ widths.  Ly$\alpha$ and
C\,IV widths are correlated with each other, however (significant at the
0.01\% level, 2-tailed). We recall that the strength of the lower-velocity
C\,IV emission is strongly correlated with the strength of
[O\,III]\,$\lambda$5007 emission (Figs. 1--3).  So part of the C\,IV
profile becomes {\em narrower} as Fe\,II(optical) increases, or as
H$\beta$ becomes {\em broader}.  So the lack of correlation between C\,IV
and H$\beta$ widths is not so surprising. This lower-velocity
C\,IV-emitting region, called the ILR, an intermediate line region
between the highest-velocity VBLR and the NLR, was related to other UV
lines in high redshift QSOs by Wills et al. (1993) and Brotherton et al.
(1994a,b) (see also, Brotherton \& Francis 1999).  The trends for high-z
QSOs are consistent with our present results for the low-redshift QSO
sample. Note that the `narrow-line' QSOs (e.g., Foltz et al. 1987,
Baldwin et al. 1988) refer to the strength of the ILR in C\,IV --
inversely related to the width of H$\beta$ that defines the NLS1s and
NLQ1s! The results of the previous section (Fig.\,3) show that much of
the Fe\,II(optical) emission has narrow widths like H$\beta$, consistent
with both species arising in the same dense gas.

The H$\beta$ widths have been suggested as an indicator of low Black Hole
mass for NLS1s (Laor, these Proceedings).  It has been suggested that
H$\beta$ and CIII] might arise from an optically-thick disk, and C\,IV
and Ly$\alpha$ may be produced in a high-ionization wind (Murray \&
Chiang 1998, Bottorff et al. 1997, K\"onigl \& Kartje 1994).  If the emission
lines are produced by different kinematic gas components, then a direct
kinematic interpretation of their FWHM is not reasonable. Note that the line
widths for Ly$\alpha$ and C\,IV are always broader than that of H$\beta$
for the NLQ1s in our sample.  This may be consistent with the suggestion
that NLQ1s tend to be lower mass objects with high accretion rates that
drive an outflow; the higher-ionization lines may be partly broadened by
radiation-pressure driven outflow.

\begin{figure}[htb]
\centerline{\psfig{figure=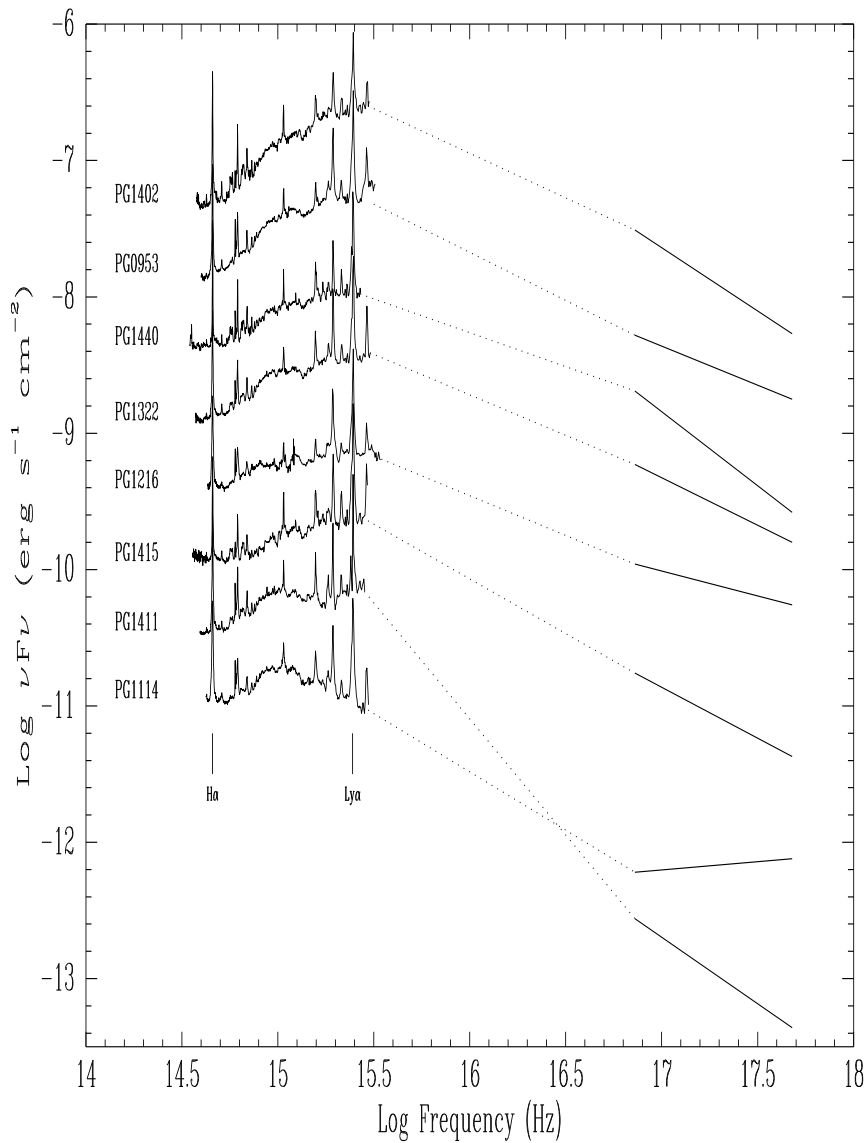,height=16truecm,width=12.0truecm,angle=0}}
\caption{
Above are Spectral Energy Distributions of QSOs.  The optical-UV spectra are equally
displaced in log\,$\nu{\rm F}_{\nu}$ in the continuum at H$\alpha$.  The H$\alpha$
and Ly$\alpha$ emission lines are marked.  The soft X-ray
spectra are ROSAT data from Laor et al. (1994, 1997), and the observational
uncertainties in these plots are very small on the scale of this figure.
The ROSAT spectra are joined by a dotted line to the UV continuum.
All spectra are corrected for absorption in our Galaxy.
}
\end{figure}

\section{The Spectral Energy Distributions}
 
We compare our well-calibrated HST-McDonald spectral energy distributions 
with the ROSAT soft X-ray spectra (H$\alpha$ to 2 keV, Fig.\,5).  The spectra
are separated by equal logarithmic increments in $\nu{\rm L}_{\nu}$ at the
continuum wavelength of H$\alpha$.
While it is difficult to illustrate with just the few spectra of Fig.\,5, 
we find some interesting trends in the whole sample:
\begin{itemize}
\item{QSOs with strong UV absorption lines have redder optical-UV continuua,
and 
either weaker or flatter soft X-ray spectra.  This is the case for the
following QSOs of our sample: 1114+445, 1411+442, 1309+355
(radio intermediate),
1001+054 (a NLQ1), and 1425+267 (radio-loud).}
\item{Relative to the optical-UV spectra, anchored at the H$\alpha$ continuum,
the soft X-ray spectra point towards the UV spectra -- either having a steeper
X-ray continuum with weaker hard X-rays, or a flatter, but stronger X-ray
spectrum.
This effect is significant for the whole sample at about the 3$\sigma$ level.}
\item{The well-known trend for radio-loud QSOs to have stronger, flatter soft
X-ray
continua is shown clearly for PG 1226+023 (3C\,273) and PG 1512+370 (4C\,37.43)
(not illustrated here).}
\end{itemize}

The NLQ1s partake of these trends (although none in our sample is radio-loud).

\section{Summary}

The NLQ1s of our sample follow the Eigenvector 1 relationships demonstrated
for the whole QSO sample in Figs.\,1 and 2.  Not only is this true for the
H$\beta$--[O\,III] region, but it extends to the UV spectra as well.
The NLQ1s therefore show narrow C\,III]\,$\lambda$1909 emission lines,
small ratios of C\,IV\,$\lambda$1549/Ly$\alpha$, larger ratios of 
N\,V\,$\lambda$1240/C\,III]\,$\lambda$1909, 
Si\,III]\,$\lambda$1892/C\,III]\,$\lambda$1909, and $\lambda$1400/Ly$\alpha$.
These trends suggest copious amounts of high-density gas in NLQ1s,
which is probably the same gas
that suppresses the ionizing-photon flux available to the extended NLR.
Strong nitrogen emission suggests that the gas may be enriched, perhaps by
starbursts from mergers that fuel the high Eddington accretion rates in NLQ1s.

The high Eddington accretion rates may drive high-ionization outflows; this
could explain the broader Ly$\alpha$ and C\,IV$\lambda$1549/Ly$\alpha$ emission
lines in NLQ1s.

The soft X-ray spectra `point at' the UV continuum.  The steep X-ray spectra of
the NLQ1s correspond to strong UV bumps in these QSOs.  However, absorption
apparently causes reddening of the optical-UV SEDs, strong UV line absorption,
and suppression of soft X-rays.

\ack

We thank our collaborators A. Laor, D. Wills, M. Brotherton, B. Wilkes and G. Ferland.
B.J.W. is very grateful to the organizing committee, and the Wilhelm und Else 
Heraeus-Stiftung, for support to attend this exciting meeting.
We thank the staff of McDonald Observatory, especially David Doss and Marian Frueh,
also Anne Kinney, Jen Christiansen and Tony Keyes of the Space Telescope Science
Institute.
The research of B.J.W. has been supported by LTSA grant NAG5-3431 and grant GO-06781 
from the Space Telescope Science Institute, which is operated by
the Association of Universities for Research in Astronomy, Inc., under
NASA contract NAS5-26555.



\end{document}